\documentstyle[12pt]{article}
\input{tcilatex}

\begin{document}

\title{A condition on the chiral symmetry breaking solution of the Dyson-Schwinger
equation in three-dimensional QED}
\author{Guo-Zhu Liu$^{1,3}\thanks{E-mail:gzliu@mail.ustc.edu.cn}$ and G.Cheng$^{2,3}\thanks{E-mail:gcheng@ustc.edu.cn}$ \\
$^{1}${\small {\it Structure Research Laboratory, }}\\
{\small {\it University of Science and Technology of China, }}\\
{\small {\it Hefei, Anhui, 230026, P.R. China }}\\
$^{2}${\small {\it CCAST (World Laboratory), P.O. Box 8730, Beijing 100080,
P.R. China }}\\
{\small {\it $^{3}$Department of Astronomy and Applied Physics, }}\\
{\small {\it University of Science and Technology of China, }}\\
{\small {\it Hefei, Anhui, 230026, P.R. China}}}
\date{13,Apr 2001}
\maketitle

\begin{abstract}
\baselineskip20pt In three-dimensional QED, which is analyzed in the 1/$N$
expansion, we obtain a sufficient and necessary condition for a nontrivial
solution of the Dyson-Schwinger equation to be chiral symmetry breaking
solution. In the derivation, a normalization condition of the Goldstone
bound state is used. It is showed that the existent analytical solutions
satisfy this condition.\newline
\end{abstract}
\newpage

\baselineskip20pt

The interest in quantum electrodynamics in two space and one time dimensions
(QED$_{3}$) is twofold. On the one hand, QED$_{3}$ itself as a quantum field
theory exhibits many interesting features, such as dynamical chiral symmetry
breaking (DCSB) \lbrack 1-5\rbrack\ and confinement \lbrack 6\rbrack , which
also exist in QCD$_{4}$ and other gauge theories. Furthermore, unlike four
dimensional QED, QED$_{3}$ has a dimensional coupling constant $e$ and
therefore is a superrenormalizable field theory, which makes it easier to be
treated. It is expected that the study on QED$_{3}$ would shed light on our
understandings of more complicated gauge theories like QCD. On the other
hand, QED$_{3}$ has been used to describe some planar condensed matter
systems, especially high temperature cuprates. It is well established that
an undoped cuprate is a 2d quantum antiferromagnet and described by the spin-%
$1/2$ Heisenberg model \lbrack 7\rbrack , which can be mapped into a theory
of massless fermions coupled to a U(1) gauge field \lbrack 8\rbrack . The
chiral symmetry breaking in this U(1) gauge theory corresponds to the
N\'{e}el ordering in the antiferromagnet \lbrack 8,9\rbrack .

One of the most interesting characteristics of QED$_{3}$ is that it may
exhibit DCSB, from which a massless fermion acquires a dynamically generated
mass. This is a nonperturbative effect and conventional perturbative
expansion approach is unable to investigate this problem. The standard
mathematical tool of \ analyzing chiral phase transition is the
Dyson-Schwinger (DS) nonlinear integral equation for the fermion
self-energy. If the DS equation has only vanishing solution, the theory is
chirally symmetric and the fermions remain massless. The case of interest to
us is when a nontrivial solution of the DS equation occurs. Generally, the
existence of a nontrivial solution of the DS equation was believed to lead
unambiguously to the existence of DCSB. However, as pointed out by Cheng and
Kuo \lbrack 10\rbrack , where they discussed this problem in quenched planar
QED$_{4}$, this is not always the case. Actually, the breaking of a global
chiral symmetry is always accompanied by a (only one in U(1) gauge theory)
massless Goldstone boson \lbrack 11\rbrack , which is a pseudoscalar bound
state composed of a fermion and an antifermion \lbrack 12\rbrack . Once the
DS equation develops a nontrivial solution, there should also be a
nontrivial solution for the corresponding Bethe-Salpeter (BS) equation
satisfied by this bound state wave function. Note that: these solutions are
not independent and can be converted into each other under a simple
transformation, and the BS wave function must satisfy an additional
normalization condition to ensure the bound state is stable \lbrack
13\rbrack . Using these two facts, we may obtain a constraint on the
nontrivial solution of the DS equation from the normalization condition of
the BS wave function. Therefore, that the DS equation has a nontrivial
solution is only a necessary but not a sufficient condition to lead to DCSB.

In the spirit of the above discussions, Cheng and Kuo \lbrack 10\rbrack\
have considered the quenched planar QED$_{4}$ and acquired a necessary and
sufficient condition for a nontrivial solution of the DS equation to be
symmetry breaking. They applied this condition to the explicit solution of
the DS equation given by Kondo $et$ $al.$ \lbrack 14\rbrack , and found
that, although DCSB occurs when there is a cutoff, once the cutoff is taken
away DCSB disappears.

In this paper, we examine this problem in the continuous QED$_{3}$ and
verify whether or not the existed nontrivial solutions of the DS equations
lead to DCSB. The Lagrangian of QED$_{3}$ in Euclidean space is

\begin{equation}
L=\sum_{i=1}^{N}\overline{\psi }_{i}(i\partial _{\mu }+eA_{\mu })\gamma
_{\mu }\psi _{i}+\frac{1}{4}F_{\mu \nu }^{2},
\end{equation}
with $N$ fermions which are four-component spinors. In this case, the 4$%
\times 4$ $\gamma _{3}$ and $\gamma _{5}$ matrices can be well constructed,
which anticommute with $\gamma _{0},$ $\gamma _{1}$, and $\gamma _{2}$. For
details about the $\gamma $ matrices, see Ref. \lbrack 2\rbrack . The
Lagrangian is invariant under the chiral transformations $\psi \rightarrow
\exp (i\theta \gamma _{3})\psi $ and $\psi \rightarrow \exp (i\omega \gamma
_{5})\psi $ because the fermions are massless. A mass term $m\overline{\psi }%
\psi ,$ no matter it is added to the Lagrangian by hand or dynamically
generated, would break these symmetries. Since the mechanism of dynamical
mass generation for fermions needs no additional Higgs particles, it is very
interesting to study DCSB in various field theories.

Power-counting arguments show that the square of the coupling constant $%
e^{2} $ has dimensions of mass, therefore QED$_{3}$ is a superrenormalizable
field theory and ultraviolet (UV) finite. However, while the UV behavior
becomes better, in the massless case, perturbative expansions in $e^{2}$
lead to infrared (IR) divergences which appear already at two loops. One way
to overcome this difficulty is using the 1/$N$ expansion \lbrack 15\rbrack .

We write the full fermion propagator as 
\begin{equation}
S^{-1}(p)=i\gamma \cdot pA(p^{2})+B(p^{2}),
\end{equation}
where $A(p^{2})$ is the wave-function renormalization and $B(p^{2})$ is the
fermion self-energy. The DS equation for the full fermion propagator \lbrack
16\rbrack\ is

\begin{equation}
S^{-1}(p)=S_{0}^{-1}(p)-e^{2}\int \frac{d^{3}q}{(2\pi )^{3}}\gamma _{\mu
}S(q)\Gamma _{\nu }(p,q)D_{\mu \nu }(p-q),
\end{equation}
where $\Gamma _{\nu }(p,q)$ is the full vertex function and the full photon
propagator is

\begin{equation}
D_{\mu \nu }(q)=\left( \delta _{\mu \nu }-\frac{q_{\mu }q_{\nu }}{q^{2}}%
\right) \frac{1}{q^{2}\lbrack 1+\Pi (q^{2})\rbrack }+\xi \frac{q_{\mu
}q_{\nu }}{q^{4}},
\end{equation}
with the vacuum polarization $\Pi (q^{2})$ related to the vacuum
polarization tensor $\Pi _{\mu \nu }(q)$ as follows:

\begin{equation}
\Pi _{\mu \nu }(q)=(q^{2}\delta _{\mu \nu }-q_{\mu }q_{\nu })\Pi (q^{2}).
\end{equation}

Substituting Eq.(2) into Eq.(3) and taking trace on both sides, we obtain
the equation for $B(p^{2})$%
\begin{equation}
B(p^{2})=-\frac{e^{2}}{4}\int \frac{d^{3}q}{(2\pi )^{3}}tr\lbrack \gamma
_{\mu }S(q)\Gamma _{\nu }(p,q)D_{\mu \nu }(p-q)\rbrack .
\end{equation}

Multiplying both sides of Eq.(3) by $\gamma \cdot p$ and then taking trace
on both sides, we obtain the equation for $A(p^{2})$%
\begin{equation}
A(p^{2})=1+\frac{e^{2}}{4p^{2}}\int \frac{d^{3}q}{(2\pi )^{3}}tr\lbrack
(i\gamma \cdot p)\gamma _{\mu }S(q)\Gamma _{\nu }(p,q)D_{\mu \nu
}(p-q)\rbrack .
\end{equation}

In this paper, in order to satisfy the Ward-Takahashi identity, the vertex
function \lbrack 17\rbrack\ is taken to be 
\begin{equation}
\Gamma _{\mu }(p,q)=\gamma _{\mu }G(p^{2},q^{2}).
\end{equation}
The one-loop vacuum polarization \lbrack 2\rbrack\ is\ 
\begin{equation}
\Pi (p^{2})=\alpha /\mid p\mid
\end{equation}
with $\alpha =Ne^{2}/8.$

Now, we can write the equation for $B(p^{2})$ as:

\begin{equation}
B(p^{2})=-\frac{e^{2}}{4}\int \frac{d^{3}q}{(2\pi )^{3}}\frac{%
B(q^{2})G(p^{2},q^{2})}{q^{2}A^{2}(q^{2})+B^{2}(q^{2})}tr\lbrack \gamma
_{\mu }\gamma _{\nu }D_{\mu \nu }(p-q)\rbrack .
\end{equation}

To look for DCSB in QED$_{3},$ the nonlinear integral equations for $%
A(p^{2}) $ and $B(p^{2})$ should be solved explicitly. It is easy to see
that there is always a trivial solution $B(p^{2})\equiv 0,$ which means
there is no mass generation of fermions and hence no DCSB. The conventional
opinion claimed that a nontrivial solution $B(p^{2})$ leads to DCSB.
However, according to the well-known Goldstone theorem \lbrack 11\rbrack ,
the spectrum of physical particles must contain one massless spin-0 particle
for each broken continuous symmetry. If the broken symmetry is a global
chiral symmetry, the corresponding Goldstone boson is a fermion-antifermion (%
$A$ and $\overline{A}$) pseudoscalar bound state. When we study DCSB, the
properties of the Goldstone boson must be taken into account. We have seen
in Ref. \lbrack 10\rbrack\ that the normalization condition of this bound
state places a nontrivial constraint on the asymptotic form of $A(p^{2})$
and $B(p^{2})$ in quenched planar QED$_{4}.$ Now, we address this constraint
in the three-dimensional QED.

This bound state is a nontrivial solution of the following BS equation in
Euclidean space \lbrack 18\rbrack : 
\begin{equation}
S_{A}^{-1}(\frac{k}{2}+p)\chi _{k}(p)S_{\overline{A}}^{-1}(\frac{k}{2}%
-p)=-e^{2}\int \frac{d^{3}q}{(2\pi )^{3}}\gamma _{\mu }\chi _{k}(q)\Gamma
_{\nu }(p,q)D_{\mu \nu }(p-q).
\end{equation}
Here, $k/2=(p_{A}+p_{\overline{A}})/2$ and $p=(p_{A}-p_{\overline{A}})$ are
the center-of-mass 3-momentum and relative 3-momentum of the bound state,
respectively. $\chi _{k}(p)$ is the bound state wave function expressed as a
4$\times 4$ matrix.

Combining Eq.(2) and Eq.(11), we have 
\begin{eqnarray}
&&\left[ i\gamma \cdot \left(\frac{k}{2}+p\right)A \left( \left(\frac{k}{2}+p \right)^{2} \right)
+B \left( \left(\frac{k}{2}+p \right)^{2} \right) \right]  \nonumber \\
&&\times \chi _{k}(p) \left[ i\gamma \cdot \left(\frac{k}{2}-p \right)A \left( \left(\frac{k}{2}-p \right)^{2} \right)-B \left( \left(\frac{k}{2}-p \right)^{2} \right)\right] \nonumber \\
&=&-e^{2}\int \frac{d^{3}q}{(2\pi )^{3}}\gamma _{\mu }\chi _{k}(q)\Gamma
_{\nu }(p,q)D_{\mu \nu }(p-q)
\end{eqnarray}

\vspace{1pt}Note that since we adopt four-component spinors, there are two
matrices which anticommute with $\gamma _{0}$, $\gamma _{1}$ and $\gamma
_{2} $. The mass term $m\overline{\psi }\psi $ breaks two chiral symmetries
simultaneously. As a result there are two Goldstone bosons respectively
coupling to $\overline{\psi }\gamma _{\mu }\gamma _{3}\psi $ and $\overline{%
\psi }\gamma _{\mu }\gamma _{5}\psi $. These Goldstone bound states both
have the quantum number $l^{p}=0^{-}$, which corresponds to the broken
generator of the chiral group and vanishing 3-momentum. Their BS amplitudes
are $\chi _{k}(p)=\chi _{03}(p^{2})\gamma _{3}$ and $\chi _{k}(p)=\chi
_{05}(p^{2})\gamma _{5}$ respectively with $k=0$. It is easy to show that $%
\chi _{03}(p^{2})$ and $\chi _{05}(p^{2})$ satisfy the same BS equation and
normalization condition, which give the same constraint on the nontrivial
solution of the DS equation. For simplicity, we take $\gamma _{5}$ as the
example and the result remains correct for $\gamma _{3}$.

Substituting $\chi _{k}(p)=\chi _{05}(p^{2})\gamma _{5}$ and $k=0$ into
Eq.(12), and taking trace on both sides, we finally obtain the BS equation
for the Goldstone boson: 
\begin{equation}
\lbrack p^{2}A^{2}(p^{2})+B^{2}(p^{2})\rbrack \chi _{05}(p^{2})=-\frac{e^{2}%
}{4}\int \frac{d^{3}q}{(2\pi )^{3}}\chi _{05}(q^{2})G(p^{2},q^{2})tr\lbrack
\gamma _{\mu }\gamma _{\nu }D_{\mu \nu }(p-q)\rbrack ,
\end{equation}
where $A(p^{2})$ and $B(p^{2})$ are solutions of Eq.(6) and Eq.(7)
respectively. In the derivation of the right-hand side of this equation, we
have used the approximation $\Gamma _{\nu }(p,q)=\gamma _{\nu
}G(p^{2},q^{2}).$

The BS equation (13) is a homogeneous integral equation and hence can not
determine the bound state completely, leaving an arbitrary multiplicative
finite constant $C.$ To eliminate this uncertainty, an additional
normalization condition of $\chi _{05}(p^{2})$ is necessary. The search for
such a normalization condition has a long history, and there are several
approaches (different but equivalent) to normalize the bound state wave
function \lbrack 13\rbrack . In this paper, for convenience, we shall
utilize the form given by Suttorp \lbrack 19\rbrack . Rewriting the
normalization condition in three dimension, we have \ 
\begin{eqnarray}
&&\int d^3p\  tr \left\{\overline{\chi}_k\left(p\right)\left[i\gamma\cdot\left(\frac{k}{2}+p\right)A\left(\left(\frac{k}{2}+p\right)^2\right)+B\left(\left(\frac{k}{2}+p\right)^2\right)\right]\right. \nonumber\\
&&\left.\times\chi_k\left(p\right)\left[i\gamma\cdot\left(\frac{k}{2}-p\right)A\left(\left(\frac{k}{2}-p\right)^2\right)-B\left(\left(\frac{k}{2}-p\right)^2\right)\right]\right\} \nonumber\\
&&=\lambda\frac{dM_B}{d\lambda}
\end{eqnarray}
where $M_{B}$ is the mass of the bound state and $\lambda =e$ is the
coupling constant. Generally, it is sufficient to require that the integral
at the left-hand side of Eq.(14) has a finite value \lbrack 19\rbrack .
Setting $k=0,$ $\chi _{k}(p)=\chi _{05}(p^{2})\gamma _{5}$ and using the
identity $\gamma _{\mu }\gamma _{5}=-\gamma _{5}\gamma _{\mu }$, the
normalization condition can be written in the simple form 
\begin{equation}
\int_{0}^{\infty }dqq^{2}\lbrack q^{2}A^{2}(q^{2})+B^{2}(q^{2})\rbrack \chi
_{05}^{2}(q^{2})=finite.
\end{equation}

Now we introduce the function 
\begin{equation}
\phi (p^{2})=\lbrack p^{2}A^{2}(p^{2})+B^{2}(p^{2})\rbrack \chi _{05}(p^{2}),
\end{equation}
then Eq.(13) becomes 
\begin{equation}
\phi (p^{2})=-\frac{e^{2}}{4}\int \frac{d^{3}q}{(2\pi )^{3}}\frac{\phi
(q^{2})}{q^{2}A^{2}(q^{2})+B^{2}(q^{2})}G(p^{2},q^{2})tr\lbrack \gamma _{\mu
}\gamma _{\nu }D_{\mu \nu }(p-q)\rbrack .
\end{equation}
If there is a solution $B(p^{2})$ of Eq.(10) (the equation for $A(p^{2})$
always has a solution), then we also have a solution of Eq.(17) as $\phi
(p^{2})=cB(p^{2}),$ with $c$ an arbitrary finite constant. According to
Eq.(16), the Goldstone boson $\chi _{05}(p^{2})$ is given by $\phi (p^{2}),$ 
$A(p^{2})$ and $B(p^{2})$. Thus, we have constructed a relationship between
two systems, one is described by DS equation and the other BS equation.
Therefore, Eq.(16) and Eq.(17) may be regarded as another form of the
Goldstone theorem. \ 

Using Eq.(16), the normalization condition Eq.(15) can be written as 
\begin{equation}
\int_{0}^{\infty }dq\frac{q^{2}\phi ^{2}(q^{2})}{%
q^{2}A^{2}(q^{2})+B^{2}(q^{2})}=finite.
\end{equation}

For chiral symmetry to be broken, the nontrivial solution of the DS equation
(10) and the BS equation (13) must exist simultaneously. Besides, as a bound
state wave function, each nontrivial solution of the BS equation satisfies a
normalization condition. We can easily turn this condition into a constraint
placed on the nontrivial solution of the DS equation, and now state a
theorem.

Theorem: The necessary and sufficient condition for a nontrivial solution $%
B(p^{2})$ to be chiral symmetry breaking solution is that it must satisfy,
together with the solution $A(p^{2})$, the condition 
\begin{equation}
\int_{0}^{\infty }dq\frac{q^{2}B^{2}(q^{2})}{q^{2}A^{2}(q^{2})+B^{2}(q^{2})}%
=finite.
\end{equation}

The proof of a similar theorem in the case of QED$_{4}$ given in Ref.
\lbrack 10\rbrack\ is still valid in the current theory although the
conditions in these two cases are different. Hence, we will not present the
proof in this paper.

It is now straightforward to apply this result to the nontrivial solution of
the DS equation. Although the approximations (Eq.(8) and Eq.(9)) that we
take in this paper simplify the DS equation significantly, it appears
impossible to obtain its completely analytical solution. To solve Eq.(10)
analytically, further approximations are needed. Under the bare vertex and
one loop vacuum-polarization approximations, Nash \lbrack 4\rbrack\
considered the leading and next-to-leading terms in the 1/$N$ expansion of
the kernels in the DS equation and showed that higher-order corrections do
not alter the nature of the symmetry breaking. This result confirms the
qualitative conclusion made earlier by Appelquist and coworkers \lbrack
3\rbrack\ that there is a finite critical number of flavors $N_{c}$ above
which the DS equation for $B(p^{2})$ has no nontrivial solutions. Nash
solved the DS equation for $B(p^{2})/A(p^{2})$ in low momentum region and
showed that $N_{c}$ is gauge-independent. For $N<N_{c}$, considering only
first order of 1/$N$ in this paper, $B(p^{2})/A(p^{2})$ has the form\ 
\begin{equation}
\frac{B(p^{2})}{A(p^{2})}=p^{-\frac{1}{2}}\sin\left(\frac{1}{2}\left[ \frac{128}{3\pi^{2}N}-1\right]^{\frac{1}{2}}\left\{\ln \left[\frac{pA(0)}{B(0)}+\delta\right]\right\}\right),
\end{equation}
in the infrared region, where $\delta $ is a phase and $B(0)/A(0)$ is
finite. In the ultraviolet region ($p\gg \alpha $), the mass function $%
B(p^{2})/A(p^{2})$ falls like 1/$p^{2}$ \lbrack 2\rbrack . This result was
obtained via the operator-product expansion approach and does not depend on
the 1/$N$ expansion \lbrack 2\rbrack .

In the current case, the condition (19) becomes 
\begin{equation}
\int_{0}^{\infty }dq\frac{q^{2}}{1+q^{2}A^{2}(q^{2})/B^{2}(q^{2})}=finite.
\end{equation}
It is easy to see that there are no singularities in the integrand. The
dangers those would cause Eq.(21) invalid come from the ultraviolet and
infrared behaviors of $B(p^{2})/A(p^{2})$. Direct calculations show that the
ultraviolet and infrared forms of $B(p^{2})/A(p^{2})$\ satisfy this
condition. Thus, at least to the lowest order in $1/N$ expansion, the
nontrivial solution of the DS equation in QED$_{3}$ does lead to DCSB and
dynamical mass generation for fermions.

\vspace{1pt}In conclusion, we have obtained a sufficient and necessary
condition for a nontrivial solution of the DS equation to be chiral symmetry
breaking solution. In the derivation of this condition, the Goldstone
theorem and the normalization condition of the Goldstone bound state wave
function play important roles. It is showed that the nontrivial solutions
given by Nash and Appelquist $et$ $al$. \lbrack 2\rbrack\ satisfy this condition. Therefore we
see that QED in four dimension (in quenched planar approximation) and three
dimension (under the approximations mentioned above) have different chiral
phase structures. The former undergoes chiral phase transition only when a
cutoff is present, but the latter exhibits DCSB in the continuum form when
the number of fermions is less than a critical value. The origin of this
difference is still unknown and subjected to the future studies.

It is claimed \lbrack 4\rbrack\ that the intrinsic scale $\alpha $ divides
QED$_{3}$ into two parts and everything in high energy beyond $\alpha $ is
rapidly damped. From this point of view, it seems reasonable to construct an
equivalence between the continuous field theory, QED$_{3}$, and the
low-energy effective theory of antiferromagnet defined on discrete lattices.
In some sense, $\alpha $ is the lattice constant of QED$_{3}$ $\lbrack
8\rbrack $.

One of us (G.L) would like to thank T.J. Du, H.J. Pan and T. Tu for helpful
discussions. G.C. is supported by the National Science Foundation in China.

\end{document}